\documentclass[]{spie}  

\usepackage{amsmath,amsfonts,amssymb}
\usepackage{graphicx}
\usepackage{euclid}
\usepackage[colorlinks=true, allcolors=blue]{hyperref}
\usepackage{placeins}

\title{Toward a universal characterization methodology for conversion gain measurement of CMOS APS: application to Euclid and SVOM}

\author[a]{Le Graët J.}
\author[a]{Secroun A.}
\author[a]{Tourneur-Silvain M.}
\author[a]{Kajfasz E.}

\author[b]{Atteia J.-L.}
\author[c]{Boulade O.}
\author[b]{Nouvel de la Flèche A.}

\author[d]{Geoffray H.}
\author[a]{Gillard W.}
\author[a]{Escoffier S.}
\author[b]{Fortin F.}
\author[a]{Fourmanoit N.}
\author[a]{Kermiche S.}
\author[b]{Valentin H.}
\author[a]{Zoubian J.}
\author[$_{\,}$]{on behalf of the \Euclid Consortium.}

\affil[a]{Aix Marseille Univ, CNRS/IN2P3, CPPM, Marseille, France}
\affil[b]{IRAP, Université de Toulouse, CNRS, CNES, UPS, 31401, Toulouse, France}
\affil[c]{CEA-IRFU, Orme des Merisiers, 91190 Gif-sur-Yvette, France.}
\affil[d]{CNES, 18 Av. Edouard Belin, 31401, Toulouse, France}

\authorinfo{Further author information: Send correspondence to Jean Le Graët\\E-mail: legraet@cppm.in2p3.fr, Telephone: +33 (0)4 91 82 72 90\\ }

\begin{document} 
\maketitle

\begin{abstract}
With the expanding integration of infrared instruments in astronomical missions, accurate per-pixel flux estimation for near-infrared hybrid detectors has become critical to the success of these missions. Based on CPPM's involvement in both SVOM/Colibri and \Euclid missions, this study introduces universally applicable methods and framework for characterizing IR hybrid detectors and decorrelating their instrinsic properties.
The characterization framework, applied to the ALFA detector and \Euclid's H2RG, not only validates the proposed methods but also points out subtle behaviors inherent to each detector.

\end{abstract}


\section{Introduction}
\label{sec:intro}

With the advent of CMOS Active Pixel Sensors (APS) technology and the continuously improving performance of infrared hybrid detectors, more space and ground missions tend to include an infrared channel, whether photometric or spectroscopic. It is the case of the European Space Agency's (ESA) \Euclid mission~\cite{Mellier-2024}, launched in 2023, and the Sino-French mission SVOM~\cite{Atteia-2022} (Space-based multi-band astronomical Variable Objects Monitor), planned for launch on the 22th of June. Both aim, in different ways, at understanding the evolution of our Universe, a field of physics of strong scientific interest for the Center for Particle Physics in Marseille (CPPM). Naturally CPPM  has become involved in both projects, in particular taking responsibility for characterizing the scientific performance of their infrared detectors.
 
With missions increasingly aiming for ambitious scientific goals, the technical requirements have similarly intensified, necessitating an unprecedented understanding of detector performance right down to the pixel level. Despite significant advances, achieving accurate pixel-level performance assessment continues to pose significant difficulties. This requires not only handling millions of pixels but also accounting for interactions between pixels—in other words, considering the correlations between the various physical effects occurring within the pixels. From the perspective of detector characterization, it has become evident that we could benefit from a universal framework that considers these various factors while remaining independent of mission-specific details.

In this paper, we take the initial steps toward establishing such a framework by standardizing the conversion gain measurement of CMOS APS. Our focus is on deriving a per pixel conversion gain that is decorrelated from nonlinearity and interpixel capacitance (IPC). The approach is validated through two series of characterization of scientific performances: \Euclid's flight H2RGs~\cite{Blank-2011}, manufactured by Teledyne and SVOM's Astronomical Large Format Array (ALFA)~\cite{Gravrand-2022}, the fruit of a collaboration between CEA/LETI and Lynred. This framework not only ensures precise conversion gain measurements but also enhances the accuracy of deriving related parameters such as read noise, dark current, and quantum efficiency, all of which rely on a precise conversion gain value. Furthermore, it enables effective comparison of detector performance, underscoring its significance and potential for broader adoption.

In the following, Sect.~\ref{sect:method} thus provides a brief review of the method used to measure the conversion gain. This method relies on our original ``nonlinear mean-variance'' method combined with the correction of IPC-related bias. Section~\ref{sect:data} lays the groundwork for our characterization framework including data selection, test bench requirements, and data processing specificities. In Sect.~\ref{sect:results}, we derive the conversion gain and IPC maps within our framework for the two different detectors: \Euclid's H2RG and ALFA are both $2\,{\rm k}\times 2\,{\rm k}$ MCT-based hybrid detectors working in the short wavelenght infrared range with cutoff at respectively 2.3 and 2.1 \micron\ and pixel pitch of respectively 18 and 15 \micron.

\section{Method for conversion gain}
\label{sect:method}
The method used in this paper to measure conversion gain has been defined and validated previously (see Le Graët et al.~\cite{LeGraet-2024} for a detailed description). It is intended to be easily applicable to any CMOS APS and to give an unbiased estimate of conversion gain, decorrelated from pixels' nonlinearity and corrected from IPC bias. The main elements of this method are recalled hereafter. It may be divided into two parts: the nonlinear mean-variance method that addresses the effect of nonlinearity on the measure and the IPC correction method that provides a simple solution to correct IPC biased gain.

\paragraph{Nonlinear mean-variance method}
The nonlinear (NL) mean-variance method has been derived to take into account the nonlinearity of the pixel response in gain measurement. It is directly adapted from the well-known mean-variance method~\cite{Mortara-1981} that uses the relation between variance and mean of the measured signal to derive the conversion gain.
%
%
%
%
%
%
The issue is that the conversion gain, generally assumed to be constant, actually demonstrates a response dependent on the integrated charge. The primary reason for this dependence is that the pixel response is inherently nonlinear, a well-known fact. Most of this nonlinearity comes from the charge-to-voltage conversion, where the pn-junction capacitance decreases as charges accumulate in the pixel. The transistors used for voltage signal amplification and buffering also contribute to this nonlinearity, although their impact is usually much smaller (less than 1\%). To address the measured gain's dependence on integrated charge, our NL mean-variance method employs a nonlinear pixel response model based on a polynomial representation, rather than  a linear pixel response model as in the classic mean-variance approach.
In a previous paper~\cite{LeGraet-2024}, we proved that using a polynomial pixel response of order 2 or 3 as follows
\begin{equation}
    S = \dfrac{1}{g}(Q+\beta Q^2) \quad \text{or} \quad S = \dfrac{1}{g}(Q+\beta Q^2+\gamma Q^3) \; ,
    \label{eq:signal_nl}
\end{equation}
leads to nonlinear mean-variance equations given by respectively
\begin{equation}
    \sigma_S^2 \approx \dfrac{1}{g}\Bar{S} + 3\beta\Bar{S}^2 + \sigma_R^2 \quad \text{or} \quad \sigma_S^2 \approx \dfrac{1}{g}\Bar{S} + 3\beta\Bar{S}^2 + g\left( 5\gamma - 2\beta^2 \right) \Bar{S}^3 + \sigma_R^2 \; .
    \label{eq:nl_ptc}
\end{equation}
In these equations, $\bar{S}$ (ADU) is the mean output signal of a pixel that has integrated a charge $Q$ (\elec), $g$ denotes the conversion gain in \elec$\, {\rm ADU}^{-1}$, $\beta$ (\elec$^{-1}$) and $\gamma$ (\elec$^{-2}$) are the nonlinearity coefficients respectively of order 2 and 3, and $\sigma_R^2$ is the readout noise. In the following sections of this paper, the method based on a second-order polynomial will be referred to as NL2, and the method based on a third-order polynomial will be referred to as NL3. Then, fitting a curve of the variance as a function of the mean with a polynomial of order 2 or 3 allows the derivation of the conversion gain as the first-order coefficient.

\paragraph{IPC bias correction method}
The second method proposes a correction of the bias created by IPC. IPC originates from the close proximity of pixels, leading to a superposition of the electric fields of adjacent pixels, which results in a parasitic capacitance between them~\cite{Moore-2003}. Thus IPC induces electrical crosstalk between close pixels, creating spatial correlations that bias the estimation of the variance of the signal. The effect of IPC on signal is typically modeled~\cite{Moore-2004} as a convolution with the 2D impulse response $h$ of a pixel
\begin{equation}
\label{eq:ipc_sig}
    S_{\rm meas} = S_{\rm true} * h \; ,
\end{equation}
where $S_{\rm meas}$ is the signal detected in a pixel and $S_{\rm true}$ is the signal that would be detected if IPC were null and $h$ were a matrix unit with a central value of one. In a previous paper~\cite{LeGraet-2024}, we proposed to use a general form of the $h$ kernel such as 
\begin{equation}
\label{eq:kernel_ipc}
h =  \begin{bmatrix} 
    \alpha_1 & \alpha_2 & \alpha_3 \\
    \alpha_4 & 1 - \sum \limits_{i=1}^8 \alpha_i & \alpha_5 \\
    \alpha_6 & \alpha_7 & \alpha_8 
    \end{bmatrix} \; ,
\end{equation}
and we demonstrated that the effect of IPC on gain estimation using the mean-variance method is given by
\begin{equation}
\label{eq:ipc_coeff}
    \begin{gathered}
    g = \hat{g} \, k \; , \\
    \mathrm{with} \; k = 1 - 2\sum \limits_i \alpha_i + \left(\sum \limits_i \alpha_i \right)^2 + \left(\sum \limits_i \alpha_i^2 \right) \; ,
    \end{gathered}
\end{equation}
where $\hat{g}$ is the IPC biased gain and $g$ the ``true" or IPC corrected gain. Consequently, a simple multiplication of the measured gain by the corrective factor $k$ will suffice to calculate an IPC corrected gain. The calculation of this factor solely requires a precise measurement of the $\alpha_i$ IPC coefficients.

The two methods introduced here have been previously validated on one of \Euclid's 16 H2RG flight detectors using data from the ground characterization campaign conducted at CPPM. Nevertheless, to apply these methods to any CMOS APS used in low-light imaging, it is essential to establish a framework that enables the construction of a consistent mean-variance curve, regardless of the detector's technology or the observation strategy of the mission using these detectors. In the following section, this framework will be described.

\section{Definition of a general framework}
\label{sect:data}

In order to make a proper use of the method just defined and apply it to both ALFA and H2RG detectors, a rigorous framework must be defined. This framework should take into account the concrete configuration of each detector, the test environment and the specifics of the data taken during their characterization at CPPM. 
Clearly, each mission has its own optimized observing strategy depending on the target it aims to observe. For instance, \Euclid shall survey $15 \, 000$\,deg$^2$ of extragalactic sky, alternating photometry and spectrometry acquisitions. Integration times for spectrometry can be as long as ten minutes. Meanwhile, SVOM/Colibri's~\cite{Basa-2022} infrared channel will offer ground follow-up observations of $\gamma$-ray bursts' afterglow, consisting of several consecutive short exposures.
Consequently detectors' operation must be adapted to the mission objectives. Table~\ref{tab:cagirevsnisp} gives an overview of the main operating configurations for H2RG and ALFA on their respective projects.
When measuring the conversion gain, using each detector in its respective operating configuration, including pixel bias and electronic gain, could introduce biases due to these differences.
To prevent such biases, it's necessary to either ensure that the operational parameters don't bias the gain measurement or incorporate standard parameter values into the framework.  During characterization, it was observed that variations in wavelength and detector temperature (within 85--$100\,{\rm K}$) do not impact gain measurement. Hence, acquisitions across different wavelength bands and temperatures will be used interchangeably. Details regarding the chosen standard values and the rationale behind  selections will be provided below.

\begin{table} [ht!]
	\begin{center}
		\begin{tabular}{|l|c|c|}

            \hline
            Mission  & SVOM/Colibri   & \Euclid   \\ \hline
            Detector type & ALFA     & H2RG \\ \hline
            Operating temperature (K) & 100  & 95     \\ \hline
            Wavelength bands  &  J, H   & J, H, Y  \\ \hline
            Integration time (s) & 60 (J band)  & 600 (spectro)    \\ \hline
            Sky background (\elec/s/px)  &  152 (J) / 1250 (H)  & 2  \\ \hline
            Saturation mode  & Full Well  & ADC  \\ \hline
            Pixel bias (mV)  & 400 & 500\\ \hline
            Growth process & Liquid phase epitaxy (LPE) & Molecular beam epitaxy (MBE) \\ \hline

		\end{tabular}
	\end{center}
	\caption 
	{ Operating configurations of SVOM/Colibri's ALFA  and \Euclid's H2RG detectors.
	\label{tab:cagirevsnisp}  
	}
\end{table}

\paragraph{Data selection}

To prevent biases arising from differences in detector operation, data selection criteria must be included in the framework. The first requirement for accurately measuring the conversion gain is to ensure that the noise is dominated by photon shot noise. For \Euclid's H2RGs, flat-field ramps taken under fluxes between 16 and $1000$\,\elec${\rm s}^{-1}$ were used. For ALFA, due to its higher readout noise, approximately 5 times higher than \Euclid's H2RG's, ramps taken under fluxes between 200 and $1000$\,\elec${\rm s}^{-1}$ were used. All the acquisitions were taken at sensor temperature between $85\,{\rm K}$ and $100\,{\rm K}$. 

As persistence~\cite{Smith-2008} is not yet included in our model of pixel response, it is essential to mitigate its effects. For fluxes higher than 200\,\elec\,${\rm s}^{-1}$, it was observed for  \Euclid's H2RG and ALFA that the effect of persistence becomes negligible when the integrated flux is greater than 10\,{\rm k}\elec. Therefore, for these ramps, the frames before reaching an integrated flux of $10\,{\rm ke}^{-}$ were excluded. For lower fluxes in the case of \Euclid, as it has been demonstrated~\cite{Secroun-2018} that for ramps of 400 frames, after the first 100 frames, the persistence is negligible, these first 100 frames were excluded. Additionally, it was decided to limit the integrated fluxes to $70 \, \%$ of the full well to avoid effects appearing near the saturation of the pixel photodiode. Finally, to prevent the measurements from being biased by outlier pixels, several masks were applied that remove overall less than $3 \, \%$ of the entire matrix, as outlined below:

\begin{itemize}
    \item disconnected pixels;
    \item pixels that are saturated early in the ramp;
    \item pixels with a high baseline to avoid ADC saturation;
    \item highly nonlinear pixels, including cosmic rays, thanks to a quality factor based on the goodness of a linear fit on the ramp~\cite{Kubik-2016}.
\end{itemize}

\paragraph{Test bench}

Although standardized data selection helps avoid biases in gain measurement, the test bench used to acquire the data can also introduce systematic errors in estimating gain. Therefore, the performance of this test bench must meet minimum requirements. CPPM's test bench, dedicated to  the characterization of infrared detectors\cite{}, has been designed to meet stringent specifications. Its performance has been thoroughly validated thanks to engineering-grade detectors, proving highly efficient in minimizing systematic errors during data analysis. Table~\ref{tab:bench} presents the main critical parameters and their specifications.

\begin{table} [ht!]
	\begin{center}
		\begin{tabular}{|l|c|}

            \hline
            FPA temperature stability & $\le$ 1\,mK     \\ \hline
            Dark background level  & $\le$ 0.001 ph/s      \\ \hline
            Flux homogeneity on FPA & $\le$ 1\%      \\ \hline
            Flux stability  & $\le$ 1\% over 45 days      \\ \hline

		\end{tabular}
	\end{center}
	\caption 
	{ CPPM's characterization test bench main performance parameters.
	\label{tab:bench}  
	}
\end{table}


\paragraph{Data processing}
In addition to the data selection criteria and the performance requirements of the test bench, the data processing methods used to generate the mean-variance curve and measure IPC coefficients need to be integral to the framework. The mean-variance curves are constructed using flat field acquisitions.

Typically, one mean-variance curve is obtained from $M$ flat-field ramps taken with the same flux, using the variance and the mean across the $M$ ramps for each pixel. To reduce the number of acquisitions required to achieve a given accuracy, it is also possible to assume both that there are no spatial correlations between pixels and, that the spatial gain variations are negligible at small scales. Consequently, the variance and the mean of the signal across a box of $N\times N$ pixels can be used to build the mean-variance curve. This method requires only two similar ramps (to eliminate fixed pattern noise) to make one measurement of the conversion gain per superpixel rather than $M$. For the ALFA detector, 637 pairs of ramps meet all our selection criteria, while 376 pairs meet the criteria for the H2RG detector. The conversion gain estimation for each superpixel is then the average of the gains measured from each pair of ramps.

After measuring the conversion gain using the NL mean-variance method with the data outlined above, correcting the IPC bias requires measuring the IPC coefficients. For both detectors, techniques based on resetting a grid of separated pixels and observing the effect on the signal of their neighbors were chosen. For the \Euclid detector, the single pixel reset (SPR) method~\cite{Seshadri-2008} was used. It consists of an acquisition under dark conditions with a reset of the full detector at a nominal bias, followed by a reset of a grid of pixels at a different bias. By observing the amount of signal detected on the neighbors of the reset pixels at a different bias, the IPC coefficient $\alpha_i$ can be measured. A detailed description of how SPR was applied to \Euclid detector may be found in Le Graet et al. (2022)~\cite{LeGraet-2022}. The advantage of this method is that, because the acquisition is under dark conditions, there will be no diffusion, and thus only the IPC will be measured. Furthermore, by changing the grid of reset pixels, it is possible to measure the IPC coefficients of every pixel. Unfortunately, for ALFA, resetting a grid of pixels at a different bias is not available. However, a grid of pixels can be continuously reset during an acquisition. Using the method defined by Finger et al.~\cite{Finger-2006}, by comparing a nominal acquisition under flux and an acquisition under the same flux with a grid of pixels continuously reset, the IPC coefficients can be measured. How this method is applied to ALFA will be detailed in a future article.

In summary, the characterization framework for conversion gain measurement that we just defined is the combination of the NL mean-variance method, the correction of the IPC bias on gain measurement, the flat fields acquired with a test bench with sufficient performance, and the data selection criteria. In the following section, the results of the methodology applied to ALFA and the \Euclid detector will be presented.

\section{Conversion gain measurement results}
\label{sect:results}

The same framework has been applied to one of \Euclid's flight H2RG and to the ALFA detector. Detailed results are presented and discussed hereafter.

\subsection{Application of nonlinear mean-variance methods}

\begin{figure}[ht!]
    \begin{minipage}[b]{0.49\textwidth}
        \vspace{0.5cm}
        \centering
        \includegraphics[width=\textwidth]{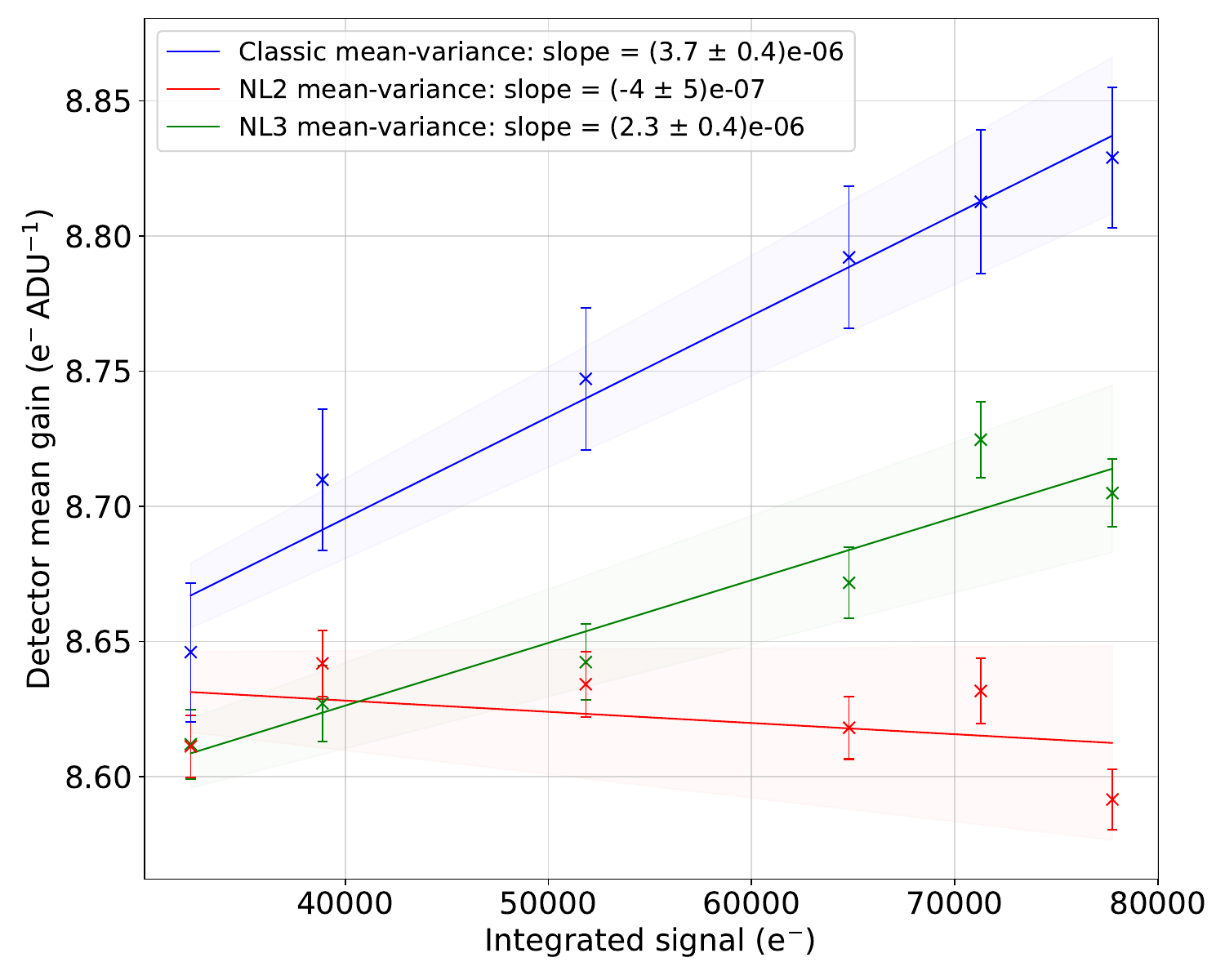}
    \end{minipage}
    \hfill 
    \begin{minipage}[b]{0.49\textwidth}
        \vspace{0.5cm}
        \centering
        \includegraphics[width=\textwidth]{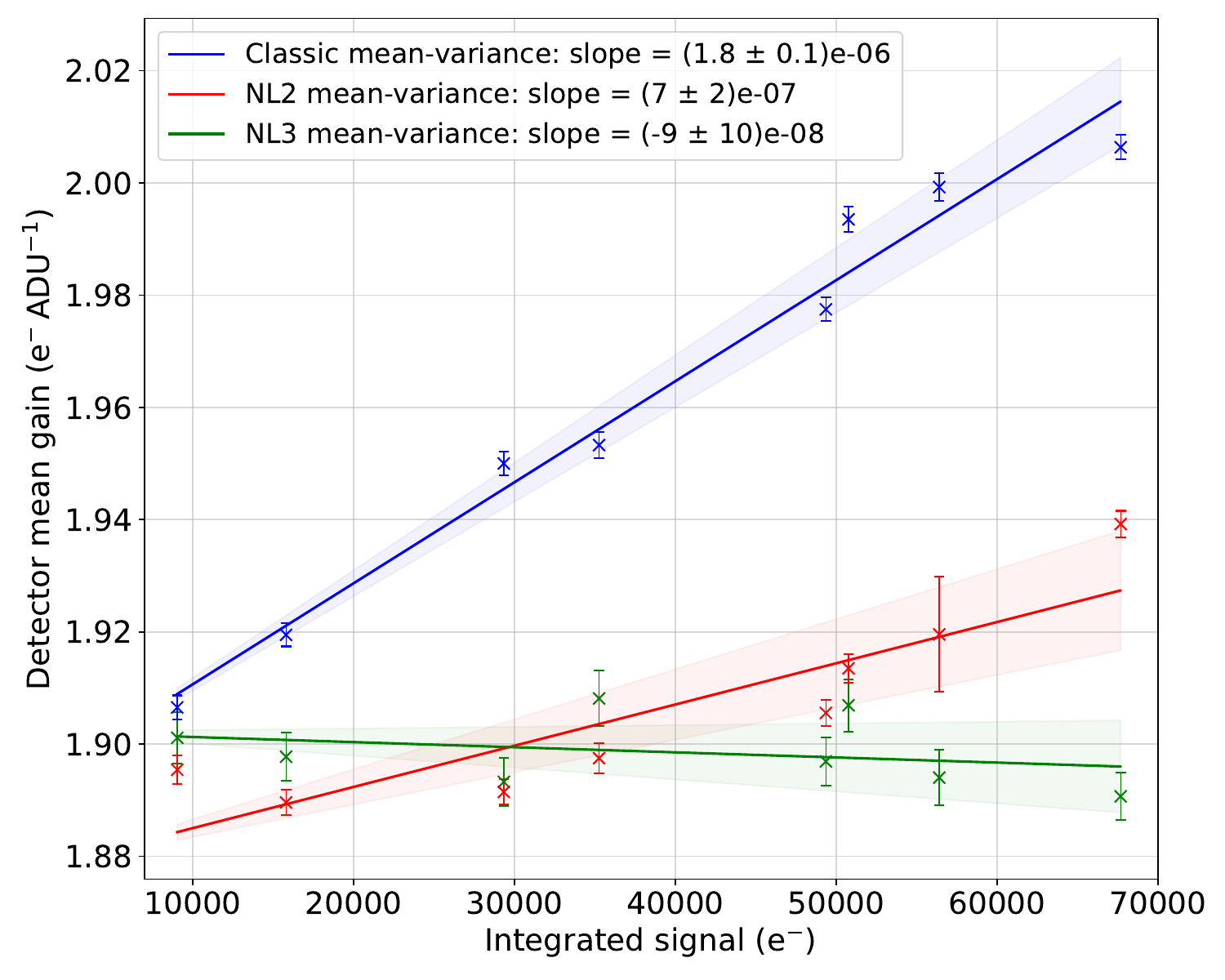}
    \end{minipage}

    \caption{
    Conversion gain, averaged over all superpixels, for ALFA (left) and \Euclid's H2RG (right) vs. integrated flux. Conversion gain obtained for all three mean-variance methods (classic, NL2 and NL3) are shown.
    \label{fig:gainvsfluence}
    }
\end{figure}

The three mean-variance methods presented in section \ref{sect:method} assume a constant gain with respect to the integrated signal (i.e., the total number of electrons accumulated by a pixel). To ascertain that the nonlinear mean-variance approach is broadly applicable, it is crucial to demonstrate that the conversion gain measured using these methods does not depend on the integrated signal. Here we have taken the opportunity to test them on two different detectors: ALFA and \Euclid's H2RG. For this purpose, the ramps selected for gain measurement were divided into subsets corresponding to equal integrated signals, and the classical, NL2, and NL3 mean-variance methods were applied to each subset. For each level of integrated signal, the gain was calculated per superpixel as the average of all corresponding measurements and then averaged across the detector. Figure~\ref{fig:gainvsfluence} shows the mean conversion gain as a function of the integrated signal (determined through LED calibration) for \Euclid's H2RG (left) and ALFA (right). The error bars include both statistical and systematic errors; the former are minimal due to averaging across the detector, while the latter arise from discrepancies in gain measurement when using ramps with identical integrated signals but different fluxes.

\FloatBarrier

As may be seen in figure \ref{fig:gainvsfluence} with the blue data,  it is obvious that, for both detectors, the conversion gain measured by the classic mean-variance method increases with the integrated signal. Specifically, for \Euclid's H2RG, the estimated gain increases by approximately $0.9 \, \%$ per $10 \, {\rm ke}^{-}$, while for ALFA, it increases by about $0.4 \, \%$ per $10 \, {\rm ke}^{-}$. Thus using this method to measure the gain will lead to a biased estimation, dependent on the integrated signal selected for measurement. Subsequent results demonstrate that both detectors exhibit comparable nonlinearity, but the impact of nonlinearity on gain measurement is less significant in the ALFA detector due to its higher conversion gain.

Regarding the NL2 mean-variance, it may be observed that \Euclid H2RG's gain remains consistent with a constant value at integrated signals below $50 \, {\rm ke}^{-}$, but increases at higher integrated signals. Conversely, ALFA's gain remains constant, within uncertainties, across all integrated signals. Additionally, the NL2 mean-variance method gives an estimate of the $\beta$ coefficient (representative of the detector nonlinear behavior): $(-4.2\pm0.1) \times 10^{-7}\,{\rm e}^{-1}$  and $(-1.3\pm0.3) \times 10^{-7}\,{\rm e}^{-1}$ respectively for \Euclid's H2RG and ALFA. Using the NL3 mean-variance, \Euclid H2RG's gain aligns with a constant value, within uncertainties, for all integrated signals, while ALFA's gain increases with integrated signal. The $\beta$ coefficients estimated for the H2RG and ALFA are $(-5.6 \pm 0.6) \times 10^{-7}\,{\rm e}^{-1}$ and $(-2 \pm 6) \times 10^{-8}\,{\rm e}^{-1}$ respectively. The $\beta$ coefficients obtained from the NL2 and NL3 mean-variance methods may be compared to those derived during the detectors' characterization. Notably, the \Euclid mission~\cite{Kubik-2014a} and the SVOM mission~\cite{delaFleche-2023} also employ models based on nonlinear pixel response as described in Eq.\eqref{eq:signal_nl}, to fit the signal ramps and correct the impact of nonlinearity on flux measurements. The $\beta$ coefficients derived from ramp nonlinearity characterization are approximately $-5 \times 10^{-7}\,{\rm e}^{-1}$ for \Euclid's H2RG and roughly $-4 \times 10^{-7}\,{\rm e}^{-1}$ for ALFA. These characterization values are comparable to those obtained from the NL2 and NL3 mean-variance methods for \Euclid's H2RG and to the one from NL2 mean-variance for ALFA.

These results demonstrate that for \Euclid's H2RG, a third-order polynomial model most accurately describes the pixel response. At high integrated signals, the increasing gain measured by the NL2 mean-variance method indicates that a second-order polynomial fails to describe the pixel response accurately. However, the similarity of the $\beta$ coefficient from the NL2 mean-variance method to the value estimated during characterization suggests that a second order polynomial may be sufficient for describing the pixel's behavior. The noted discrepancy at high integrated signal levels could be due to variations in the data related to the length of the ramp. Specifically, anomalies at the beginning and end of an acquisition significantly affect shorter ramps, such as those used for the three high-signal points (less than 100 frames). Future frameworks should incorporate ramp length to mitigate these effects. For the ALFA detector, the suitability of the second-order polynomial to describe the mean-variance curve is certain. The observation that the gain from the NL3 mean-variance increases across all integrated signals, coupled with a $\beta$ coefficient significantly different from the one obtained during characterization, underscores the model’s inadequacy in describing the pixel response. This discrepancy may originate from unaccounted effects such as persistence or diffusion, affecting the mean-variance curve. Nonetheless, the application of a nonlinear mean-variance method within a coherent framework enables the measurement of a constant conversion gain, effectively correcting for the nonlinearity of the pixel response. The flexibility to employ either the NL2 or NL3 variant allows for tailored adaptations to the distinct behaviors of the detectors.

\subsection{Correction of IPC bias on gain measurement}

\begin{figure} [ht!]
	\begin{center}
		\begin{tabular}{c} 
			\includegraphics[width=.95\columnwidth, angle=0,  trim={0 0 0 6pt}, clip]{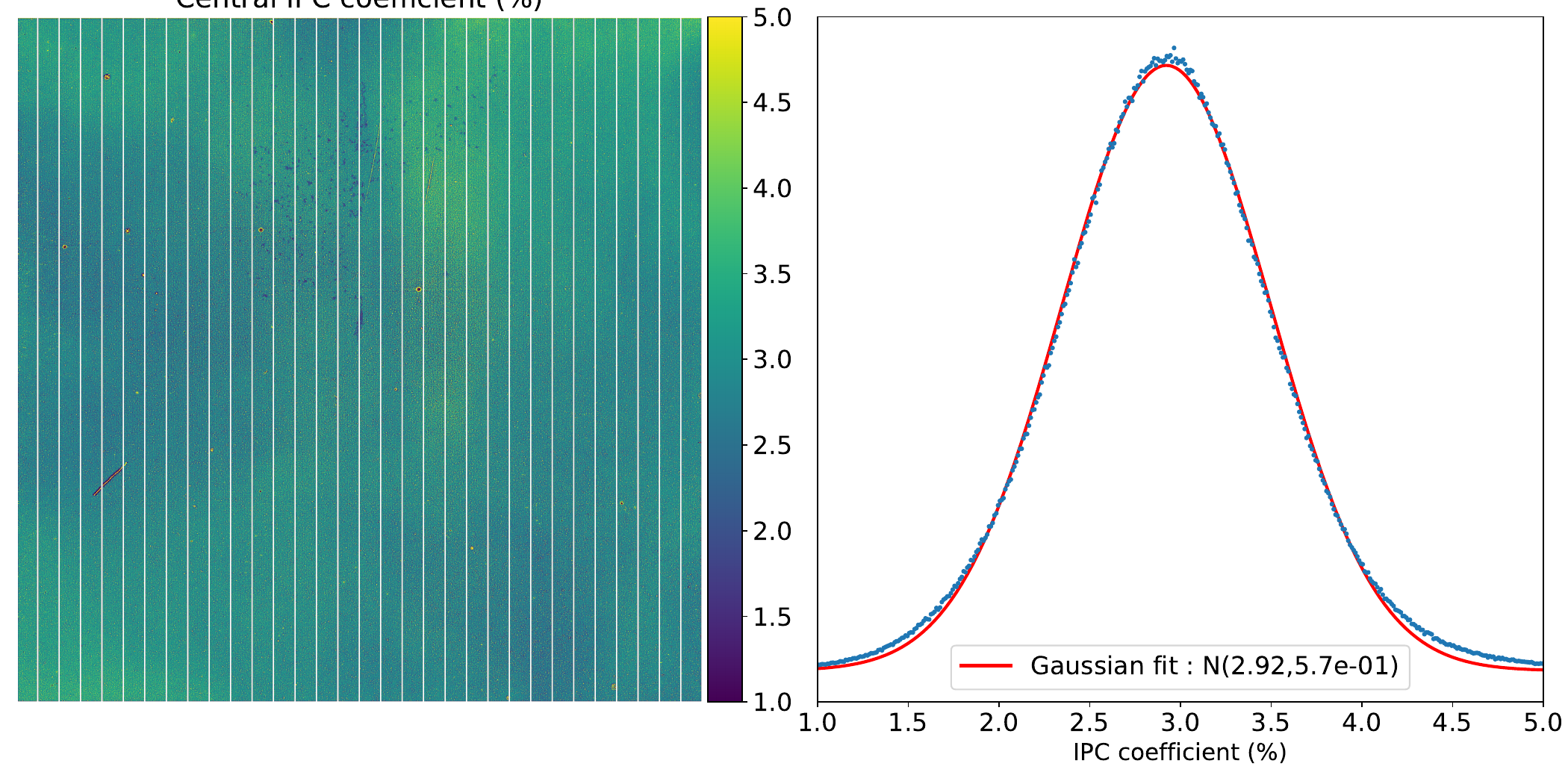}
		\end{tabular}
	\end{center}
	\caption 
	{ Total IPC (\%) for SVOM/Colibri's ALFA detector: map (left) and histogram (right).
	\label{fig:ipc_alfa}  
	}
\end{figure}

\begin{figure} [ht!]
	\begin{center}
		\begin{tabular}{c} 
			\includegraphics[width=.95\columnwidth, angle=0]{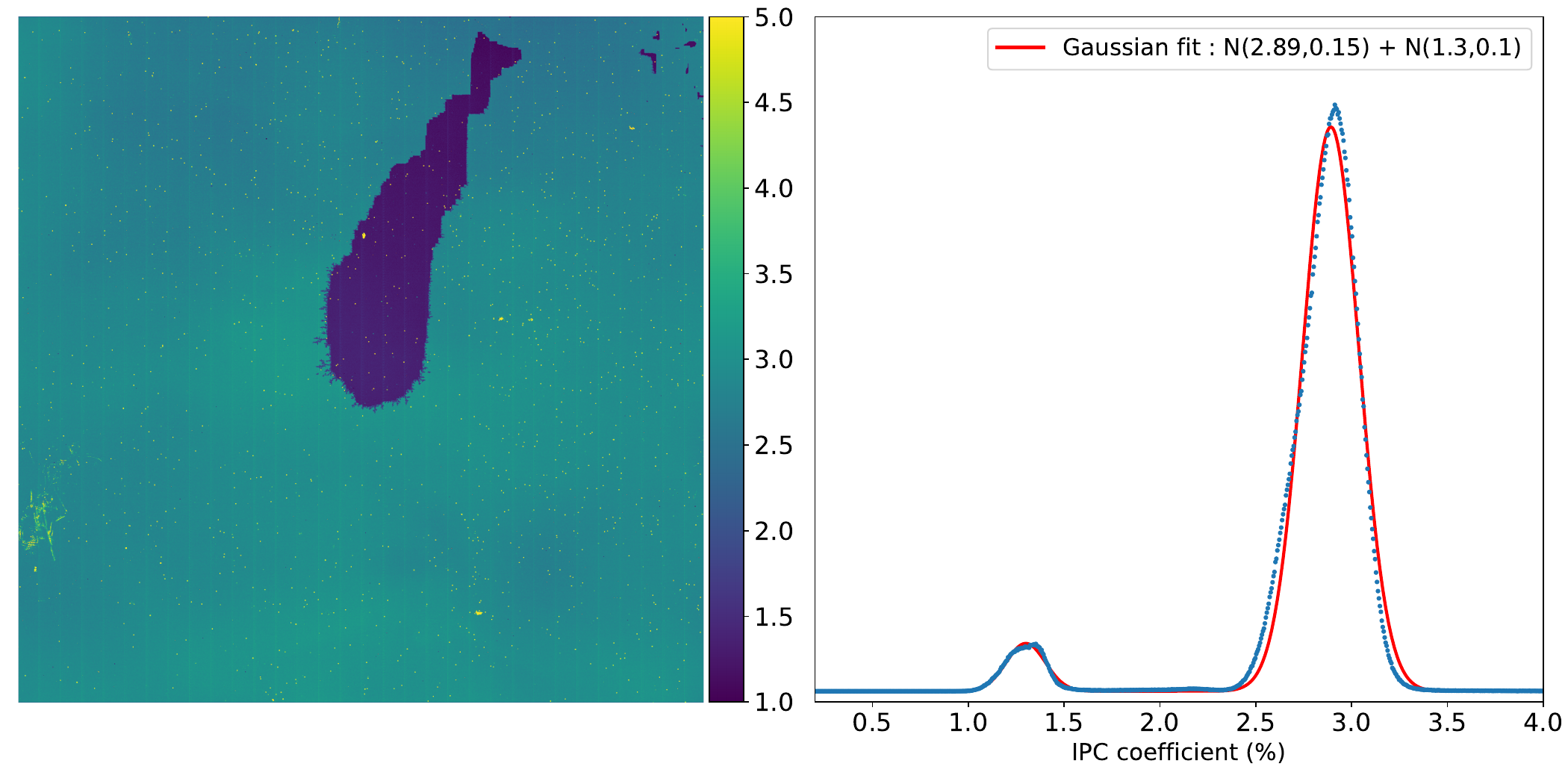}
		\end{tabular}
	\end{center}
	\caption 
	{ Total IPC (\%) for \Euclid's H2RG detector: map (left) and histogram (right).
	\label{fig:ipc_euclid}  
	}
\end{figure}

To apply the correction of IPC bias on gain measurement using the methods previously explained, the initial step is to measure the IPC coefficients for each pixel. As mentioned in Sect.~\ref{sect:data}, two distinct techniques were used for each detector, the SPR technique for \Euclid's H2RG, and the method described by Finger~\cite{Finger-2006} for ALFA. For both methods, it was decided to limit the IPC kernel to a $3 \times 3$ size as IPC is not detectable beyond this range. The eight $\alpha_i$ IPC coefficients of \Euclid's H2RG were presented in a previous publication~\cite{LeGraet-2022}, and those of ALFA will be discussed in an upcoming paper. Nevertheless maps of the total IPC (sum of the $\alpha_i$ coefficients) for \Euclid's H2RG and ALFA matrices are shown in Fig.~\ref{fig:ipc_euclid} and Fig.~\ref{fig:ipc_alfa} respectively. Due to limitations in ALFA's readout mode capabilities, it is impossible to maintain the first column of each readout channel under reset. Consequently, the corresponding IPC coefficients have not been measured. Furthermore, Table~\ref{tab:IPCcoeff} presents the median values and statistical uncertainties of the total IPC for both detectors. 

\begin{table} [ht!]
	\begin{center}
		\begin{tabular}{|l|c|c|}

            \hline
            Detector    & ALFA      & H2RG          \\ \hline
            Median IPC (\%)  & 2.92   & 2.87    \\ \hline
            Uncertainty (\%) & ± 0.12  & ± 0.01  \\ \hline
            
		\end{tabular}
	\end{center}
	\caption 
	{ Total IPC and uncertainties for SVOM/Colibri's ALFA and \Euclid's H2RG.
	\label{tab:IPCcoeff}  
	}
\end{table} 

The median total IPC is very similar for both detectors, even though ALFA pixels are closer (15\,\micron) than \Euclid H2RG's (18\,\micron). Typically, closer pixel spacing increases IPC, as demonstrated by TELEDYNE's H4RG detectors~\cite{Mosby-2020}. This suggests that LYNRED's strategies to minimize IPC have been effective. However, excluding the dark blue zone of  \Euclid H2RG's map (to be discussed later), IPC is significantly more uniform for \Euclid's H2RG (within $\pm10\, \%$ at $2\sigma$) than for ALFA (within $\pm40 \, \%$ at $2\sigma$). Factors such as the spacing and the size of the indium bumps may influence this uniformity. Nevertheless, these substantial spatial variations underline the necessity of measuring IPC on a per-pixel basis, since using an average value could introduce biases of about $10 \, \%$ for \Euclid's H2RG and $40 \, \%$ for ALFA. These biases could then propagate to the corrections made for IPC bias in both gain measurements and PSF size estimation. Observations from Fig.~\ref{fig:ipc_euclid} reveal two distinct regions: the center (dark blue) and the surrounding areas (green blue). The dark blue region has been identified as an epoxy void area, where the epoxy between the sensitive layer and the silicon multiplexer is missing. In this region, referred to hereafter as the ``void region'', IPC is more than twice as low as in the rest of the detector that will be designated as the ``epoxy region.'' This discrepancy, previously noted by Brown~\cite{Brown-2006}, is attributed to the epoxy's dielectric constant being approximately four times higher than that of air. Finally, uncertainties associated with IPC measurements suggest that the SPR method provides a more precise determination of IPC compared to the Finger method, primarily because measurements using the Finger method are constrained by photon shot noise.

\begin{figure}[ht!]
    \begin{minipage}[b]{0.49\textwidth}
        \vspace{0.5cm}
        \centering
        \includegraphics[width=\textwidth]{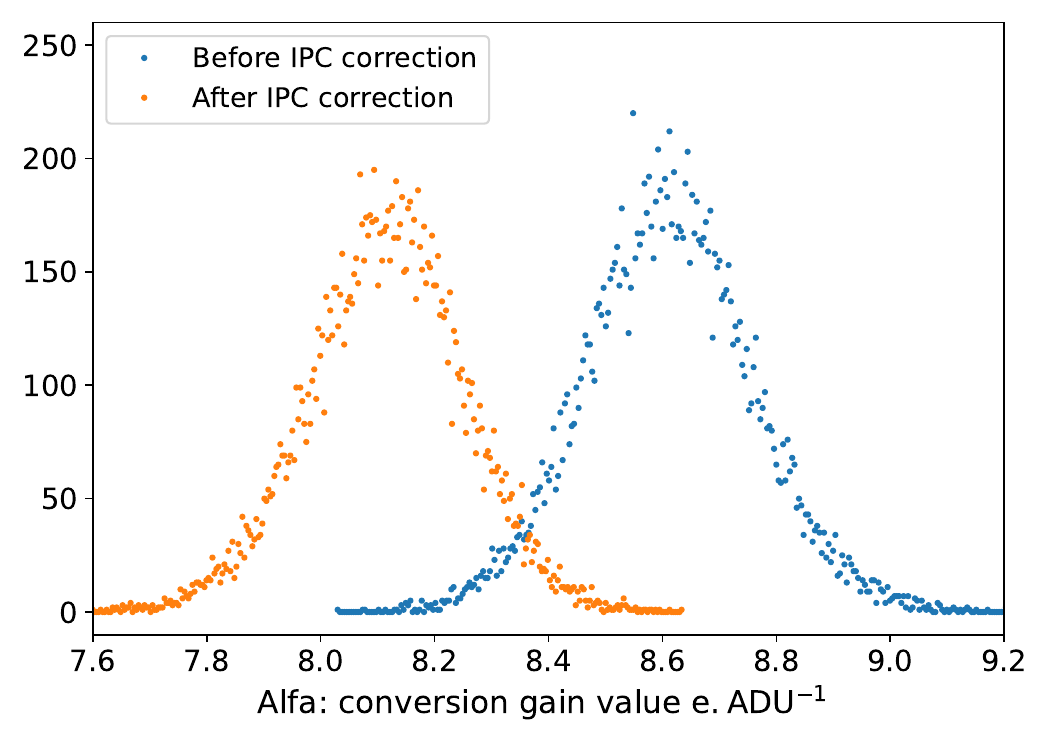}
    \end{minipage}
    \hfill 
    \begin{minipage}[b]{0.49\textwidth}
        \vspace{0.5cm}
        \centering
        \includegraphics[width=\textwidth]{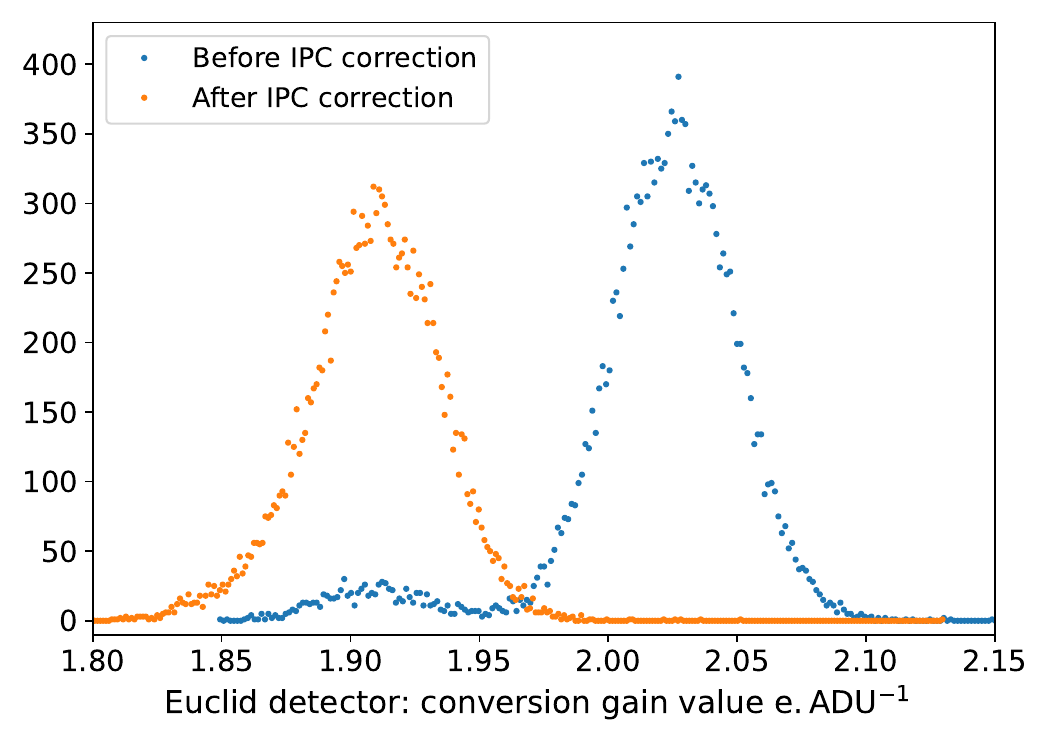}
    \end{minipage}

    \caption{
    Histogram of the conversion gain across all superpixels for ALFA (left) and the \Euclid detector (right), before and after correction of the IPC bias.
    \label{fig:gain_ipc}
    }
\end{figure}

Thanks to the measurements of IPC for each pixel of the detectors, it is now possible to calculate the corrective factor as defined in Eq.~\eqref{eq:ipc_coeff}. Given that superpixels of size $16\times 16$ were used, the corrective factors were averaged within each superpixel. These averaged values were subsequently applied to the gains derived using the NL3 mean-variance method for \Euclid's H2RG and the NL2 mean-variance method for ALFA . The gain estimation is the mean of the gains derived from each pair of ramps selected according the criteria outlined in Sect.\ref{sect:data}. The histograms representing the superpixel's  conversion gains of ALFA (left) and \Euclid's H2RG (right) detectors, both before and after correction, are illustrated in Fig.~\ref{fig:gain_ipc}. The main outcome from these figures is that this method corrects a bias in the gain measurement by approximately $6 \, \%$ for ALFA and $5 \, \%$ for \Euclid's H2RG. The histogram of \Euclid's H2RG before IPC correction reveals a minor peak at $1.9 \, {\rm e}^{-}{\rm ADU}^{-1}$, corresponding to the previously identified void region. Post-correction, the disparity between this peak and that corresponding to the epoxy region has significantly decreased, from $5.5 \, \%$ to $2.5 \, \%$. However, a residual difference between these two regions persists, suggesting that the epoxy void also influences the gain of the pixels. Except for this void region, the effect of correcting IPC bias on gain is very similar for both detectors. 

\FloatBarrier
\subsection{Usefullness of per superpixel unbiased conversion gain}

\begin{figure} [ht!]
	\begin{center}
		\begin{tabular}{c} 
			\includegraphics[width=.95\columnwidth, angle=0]{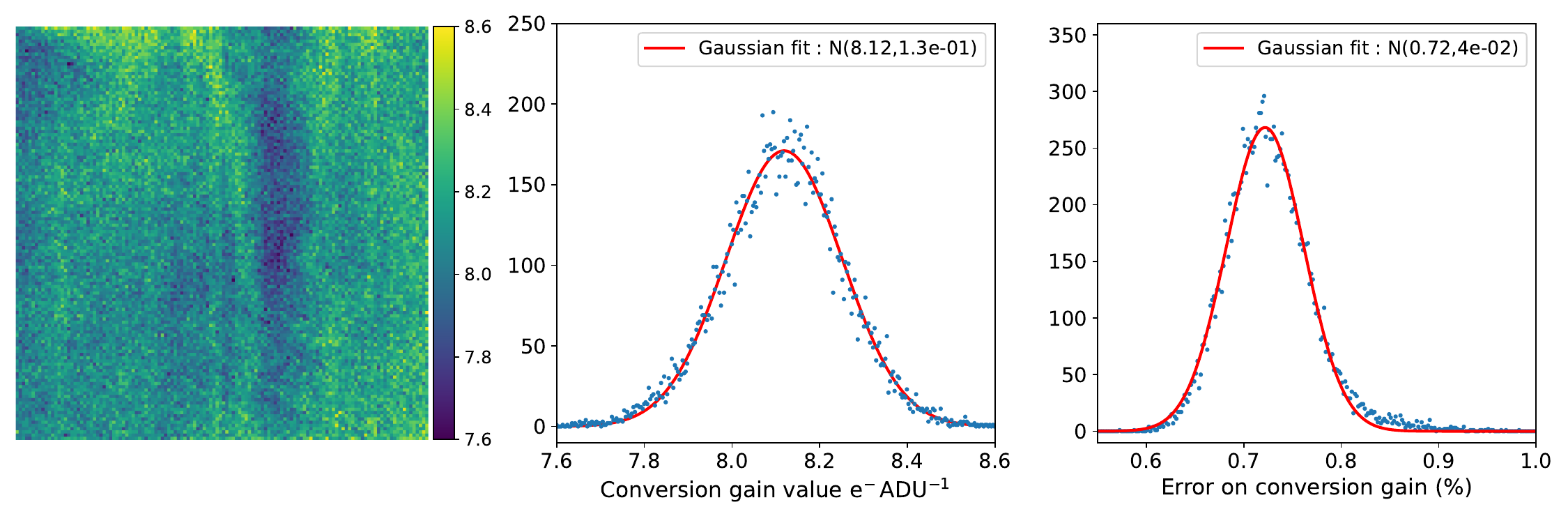}
		\end{tabular}
	\end{center}
	\caption 
	{ Spatial variations of unbiased conversion gain (\elec/ADU): map (left), histogram (middle) and error (right) for SVOM/Colibri’s ALFA
	\label{fig:gain_alfa}  
	}
\end{figure}

\begin{figure} [ht!]
	\begin{center}
		\begin{tabular}{c} 
			\includegraphics[width=.95\columnwidth, angle=0]{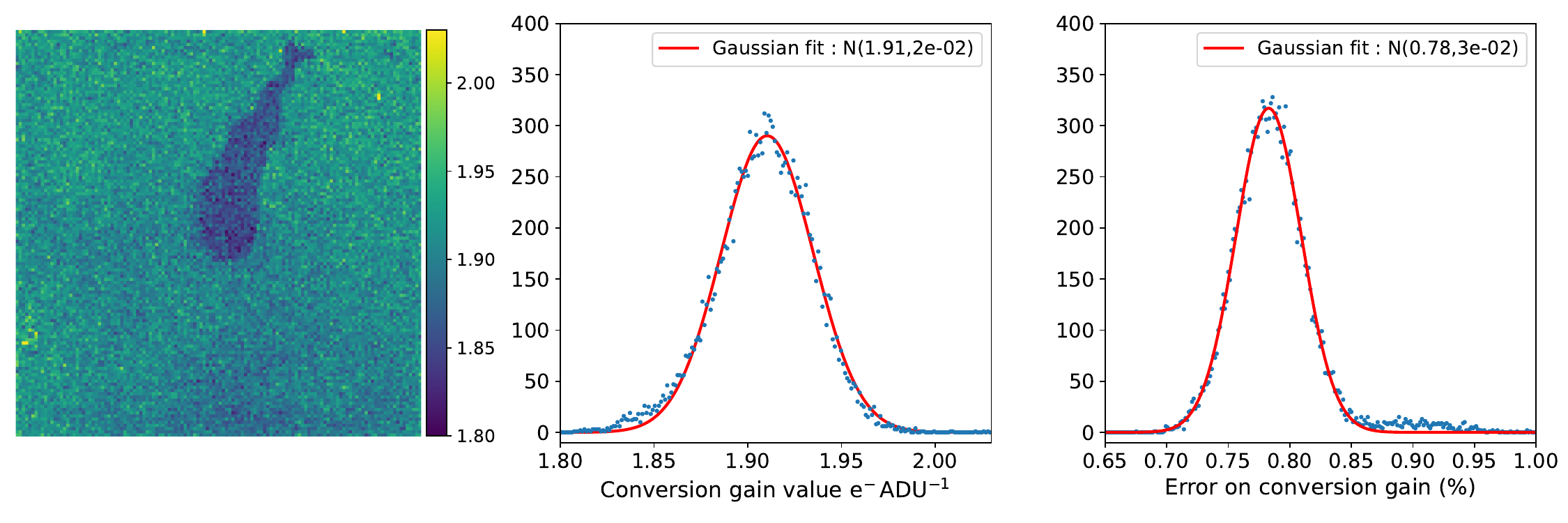}
		\end{tabular}
	\end{center}
	\caption 
	{ Spatial variations of unbiased conversion gain (\elec/ADU): map (left), histogram (middle) and error (right) for \Euclid's H2RG.
	\label{fig:gain_euclid}  
	}
\end{figure}

The methods and framework previously described result in the conversion gain maps, superpixel histograms and error histograms displayed in Fig.~\ref{fig:gain_alfa} for ALFA and Fig.~\ref{fig:gain_euclid} for \Euclid's H2RG. The error associated with the conversion gain estimation is statistically calculated using the standard deviation $\sigma_{\rm g}$ across all measured gains for each superpixel, defined as: ${\rm err} = \sigma_{\rm g} / \sqrt{M}$, where $M$ represents the number of pairs of ramps used. For ALFA, the mean gain is about $8.12\pm0.06 \, {\rm e}^{-}{\rm ADU}^{-1}  $ while it is $1.91\pm0.02 \, {\rm e}^{-}{\rm ADU}^{-1}$ for  \Euclid's H2RG. This corresponds to node capacitances of approximately $60 \, fF$ for ALFA, and $30 \, fF$ for \Euclid's H2RG. Previous analysis, conducted under settings akin to those applied in our measurements, using classic mean-variance with IPC correction, estimated ALFA's conversion gain~\cite{delaFleche-2023} as approximately $10 \, {\rm e}^{-}{\rm ADU}^{-1}$, revealing a discrepancy of $20\, \%$. This difference emphasizes the requirement to use a coherent framework to avoid discrepancies between measurements of the same parameter. For both detectors, the use of $16\times16$ superpixels achieves a measurement accuracy of gain better than $1 \, \%$, meeting the objectives outlined in this study. Obviously, to increase the resolution, more statistics are required. For \Euclid's H2RG, as previously mentioned, the conversion gain in the void region is smaller than in the epoxy region by about $2.5 \, \%$. Excluding this region, the conversion gain of  \Euclid's H2RG shows greater uniformity (within $\pm 2 \%$ at $2 \sigma$) compared to ALFA (within $\pm 3.2 \%$ at $ 2 \sigma$), as observed with the IPC coefficients. Such differences likely originate from the distinct manufacturing processes of LYNRED and TELEDYNE. These spatial variations highlight the necessity of measuring conversion gain at least on a per-superpixel basis to eliminate biases in subsequent gain-dependent measurements. Lastly, the absence of correlation observed between the conversion gain maps and the persistence maps for each detector demonstrates that the persistence mitigation strategies are effective.

Conversion gain is a fundamental parameter measured early during detector characterization because nearly all other critical parameters, such as quantum efficiency (QE), dark current, and readout noise, depend on a known gain value for their measurement. In the following discussion, we will demonstrate how a precise measurement of gain, either per pixel or per superpixel, using a coherent framework enables accurate determination of QE and allows comparison of QE across different detectors.
The standard method to measure QE involves comparing the pixel output in ${\rm e.s}^{-1}$ of the detector with that of a calibrated photodiode. However, to convert the pixel flux from ADU to \elec, it is necessary to apply a conversion gain. Typically, a mean gain value is applied across the entire detector. This approach may cause spatial variations in QE to overlap with variations from conversion gain. Nevertheless, by using the conversion gain measured per superpixel, as described previously, one can distinguish the effects of gain and QE, thereby achieving a more precise measurement of the QE.

\begin{figure} [ht!]
	\begin{center}
		\begin{tabular}{c} 
			\includegraphics[width=.85\columnwidth, angle=0]{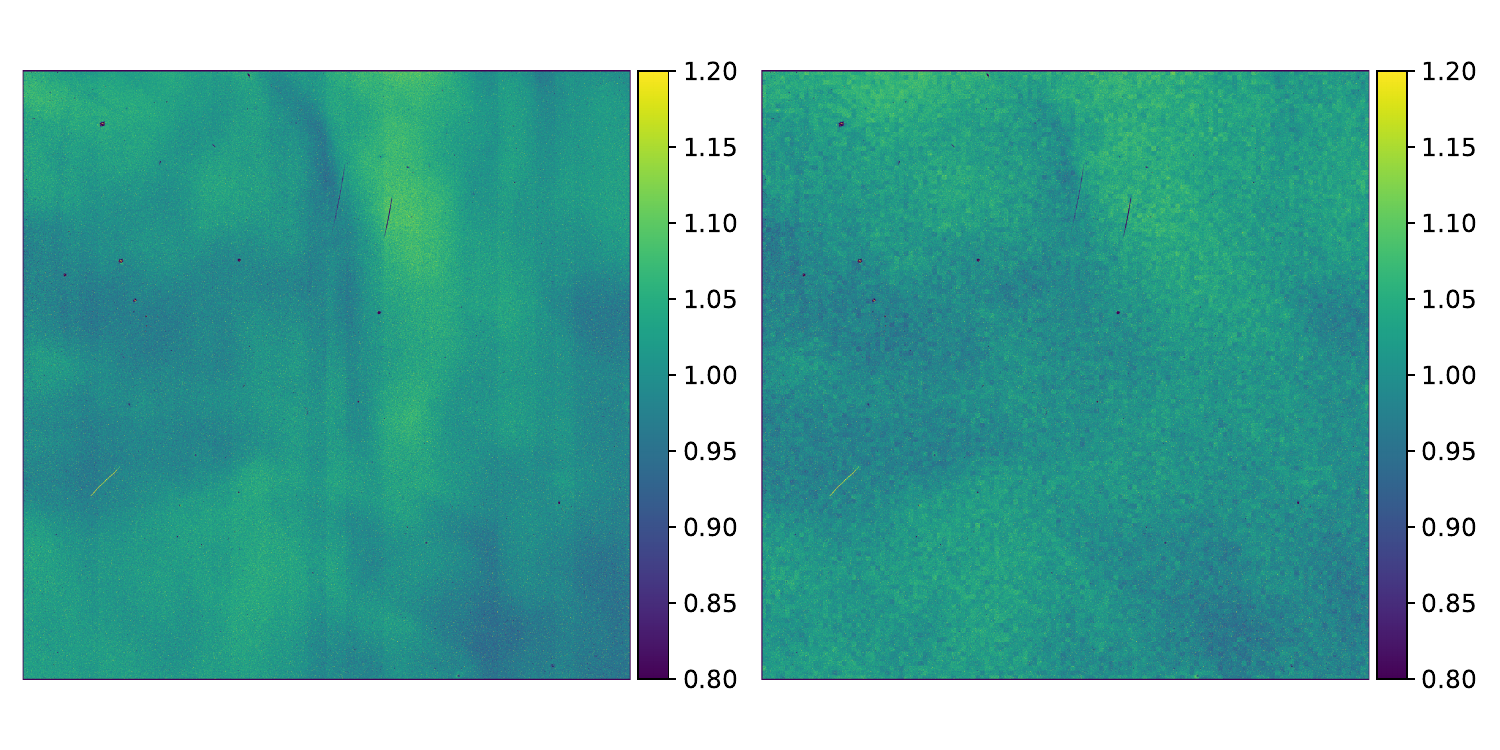}
		\end{tabular}
	\end{center}
	\caption 
 	{ Normalized relative QE of ALFA measured using mean conversion gain (left) and per superpixel conversion gain(right).
	\label{fig:flat_alfa}  
	}
\end{figure}

\begin{figure} [ht!]
	\begin{center}
		\begin{tabular}{c} 
			\includegraphics[width=.85\columnwidth, angle=0]{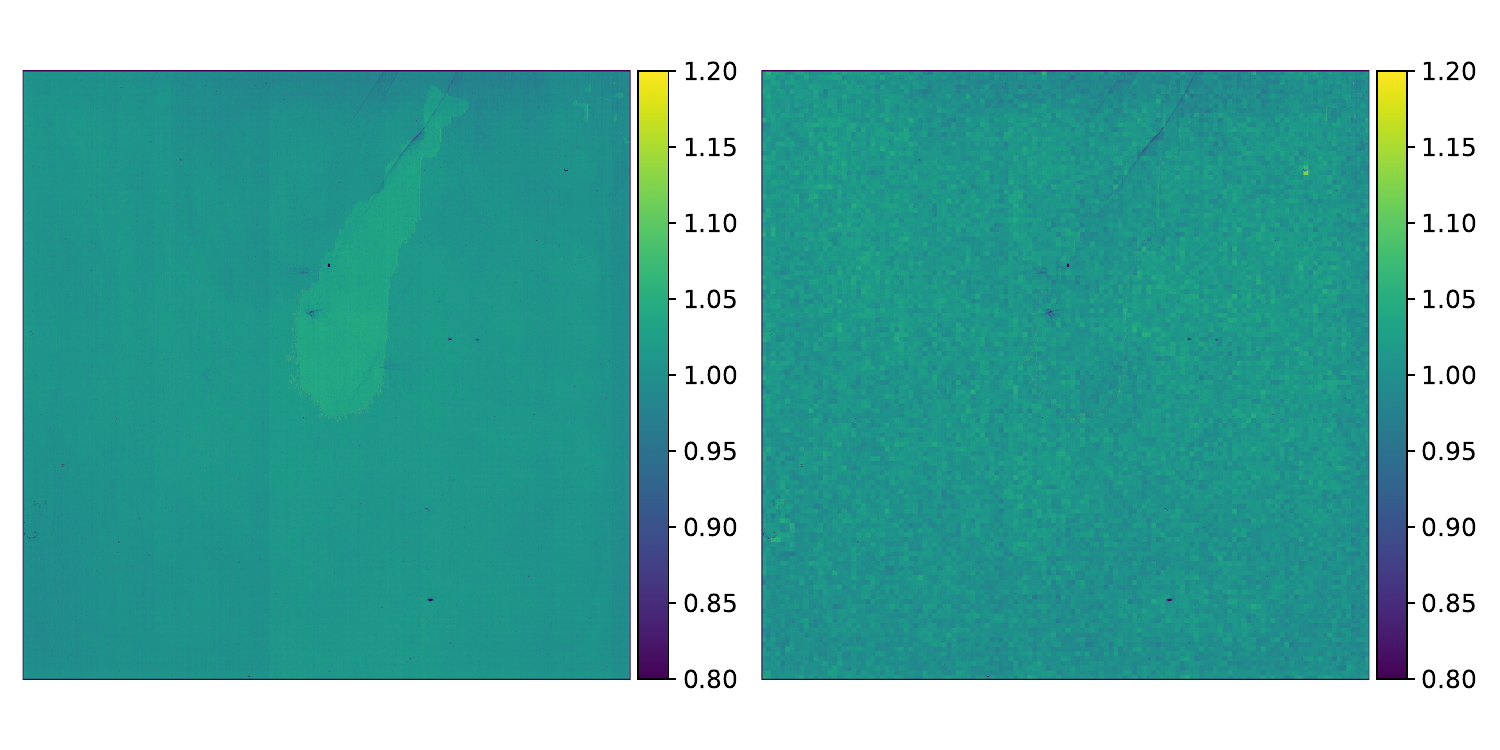}
		\end{tabular}
	\end{center}
	\caption 
 	{ Normalized relative QE of \Euclid's H2RG measured using mean conversion gain (left) and per superpixel conversion gain(right).
	\label{fig:flat_euclid}  
	}
\end{figure}

Since the test bench at CPPM does not include a calibrated photodiode, an absolute measurement of the QE for both detectors is impossible. Nevertheless, the excellent homogeneity of the flat field, better than $1\%$ at CPPM, allows relative QE measurements. To evaluate the efficiency of using a mean gain versus a per superpixel unbiased gain, one flat field acquisition per detector with fluxes of some $700 \, {\rm photons} \, {\rm s}^{-1}$ will be used. For each acquisition, the flux in ${\rm ADU} \, {\rm s}^{-1}$ will be calculated using correlated double sampling. Then, either a spatial mean or the previously measured per superpixel conversion gain will be used to convert these values from ${\rm ADU \, s}^{-1}$ to ${\rm e}^{-}{s}^{-1}$. The results will then be normalized by the spatial mean to derive a relative QE measurement. Figure~\ref{fig:flat_alfa} for ALFA and \ref{fig:flat_euclid} for \Euclid's H2RG illustrate the normalized relative QE maps obtained using mean gain (left) and per superpixel gain (right).

For both detectors, the maps appear significantly flatter when using per superpixel gain; the readout channels are no longer discernible, and regions with substantial deviations from the spatial mean are attenuated. For \Euclid's H2RG, spatial features are almost undetectable, indicating a remarkably uniform QE (and consequently, a uniform sensitive layer). However, in the ALFA detector, some areas still exhibit responses $10 \, \%$ above the spatial mean, suggesting less uniformity in the sensitive layer compared to the H2RG detectors. This difference in uniformity might come from minor instabilities during the fabrication of the sensitive layer or the differing growth techniques—LPE for ALFA versus MBE for H2RG.  It is important to note that the H2RG detectors are flight models, obviously more mature than ALFA, the first prototype of very low flux $2{\rm k}\times2{\rm k}$ NIR detector developed at CEA and Lynred.

\FloatBarrier

\section{Conclusion}
This study has successfully validated the methodologies and framework developed for characterizing the conversion gain of CMOS APS detectors, specifically using the infrared detectors of the \Euclid and SVOM missions. By employing nonlinear mean-variance methods, we have demonstrated that conversion gain can be measured as a constant across various levels of signal integration for both detectors, thereby underscoring the robustness of our methods in addressing diverse detector behaviors

Our integrated framework, in addition to methodologies that efficiently decorrelate IPC and nonlinearity of the pixel response from gain measurement, significantly reduces biases in gain estimation. By applying this approach, we have systematically eliminated biases associated with correlations of approximately $7 \, \%$. Furthermore, the development of a stringent framework that incorporates robust data selection criteria—such as excluding frames with integrated signals below $10 {\rm ke}^{-}$ to mitigate persistence and limiting measurements to $70 \, \%$ of full well capacity to avoid nonlinear effects at saturation—ensures that our gain measurements are unaffected by adverse environmental conditions. This rigorous approach has corrected previously significant discrepancies, particularly a 20\% error in previously reported gain measurements for the ALFA detector.

Moreover, the creation of accurate gain maps allows for precise measurements of pixel response parameters that are decorrelated from conversion gain. For example, it enables a precise measure of quantum efficiency (QE). By applying this framework to ALFA and  \Euclid's H2RG, we have observed significant differences in the spatial uniformity of the QE of the two detectors. The H2RG detectors used in the \Euclid mission, produced using MBE, exhibited greater uniformity, indicating a potentially more controlled manufacturing environment compared to the LPE used for ALFA. This analysis not only allows us to compare the quality of the sensitive layer but also to understand the impact of the different techniques used to produce the detectors on their performance.

In conclusion, the successful application of our characterization framework to multiple detector types not only ensures the accuracy of fundamental detector parameters but also provides a detailed evaluation of the fabrication processes and operational efficiencies across different detectors. As such, it serves as a valuable tool for advancing the field of detector technology and improving the data reliability of space and ground-based astronomical missions.

\acknowledgments 
 
This work was developed within the frame of a CNES-CNRS funded Phd thesis. 

\AckEC

\bibliography{main} 

\begin{thebibliography}{10}

\bibitem{Mellier-2024}
Mellier, Y., Abdurro'uf, and Barroso, J. A.~A., ``Euclid. {{I}}. {{Overview}}
  of the {{Euclid}} mission,'' (May 2024).

\bibitem{Atteia-2022}
Atteia, J.~L., Cordier, B., and Wei, J., ``The {{SVOM}} mission,'' {\em
  International Journal of Modern Physics D}~{\bf 31},  2230008 (Apr. 2022).

\bibitem{Blank-2011}
{Blank}, R., {Anglin}, S., and {Beletic}, J.~W., ``{The HxRG Family of High
  Performance Image Sensors for Astronomy},'' in [{\em Solar Polarization
  6}{\nolinebreak\hspace{0.1em}]},  {Kuhn}, J.~R., {Harrington}, D.~M., and
  {Lin}, H., eds., {\em Astronomical Society of the Pacific Conference Series}
  {\bf 437},  383 (Apr. 2011).

\bibitem{Gravrand-2022}
Gravrand, O., Lobre, C., and Santailler, J.-L., ``Fabrication and
  characterization of a high performance {{NIR}} 2kx2k {{MCT}} array at {{CEA}}
  and {{Lynred}} for astronomy applications,'' {\em Proceedings of SPIE - The
  International Society for Optical Engineering}~{\bf 12107},  1210706 (May
  2022).

\bibitem{LeGraet-2024}
Le~Graët, J. and Secroun, A., ``Euclid: Methodology for derivation of
  ipc-corrected conversion gain for nonlinear cmos aps,'' submitted (2024).

\bibitem{Mortara-1981}
{Mortara}, L. and {Fowler}, A., ``{Evaluations of Charge-Coupled Device / CCD /
  Performance for Astronomical Use},'' in [{\em Society of Photo-Optical
  Instrumentation Engineers (SPIE) Conference
  Series}{\nolinebreak\hspace{0.1em}]},   {\bf 290},  28 (1981).

\bibitem{Moore-2003}
Moore, A.~C., Ninkov, Z., and Burley, G.~S., ``Operation and test of hybridized
  silicon p-i-n arrays using open-source array control hardware and software,''
  {\em Proceedings of SPIE - The International Society for Optical
  Engineering}~{\bf 5017},  240--253, SPIE (May 2003).

\bibitem{Moore-2004}
{Moore}, A.~C., {Ninkov}, Z., and {Forrest}, W.~J., ``{Interpixel capacitance
  in nondestructive focal plane arrays},'' {\em Society of Photo-Optical
  Instrumentation Engineers (SPIE) Conference Series} {\bf 5167},  204--215
  (Jan. 2004).

\bibitem{Basa-2022}
Basa, S., Lee, W.~H., and Dolon, F., ``{{COLIBRI}}, a wide-field 1.3 m robotic
  telescope dedicated to the transient sky,'' {\em Proceedings of SPIE - The
  International Society for Optical Engineering}~{\bf 12182},  121821S (Aug.
  2022).

\bibitem{Smith-2008}
Smith, R.~M., Zavodny, M., Rahmer, G., and Bonati, M., ``A theory for image
  persistence in {{HgCdTe}} photodiodes,'' {\em Proceedings of SPIE - The
  International Society for Optical Engineering}~{\bf 7021},  70210J (July
  2008).

\bibitem{Secroun-2018}
Secroun, A., Cl{\'e}mens, J.-C., and Ealet, A., ``Euclid flight {{H2RG IR}}
  detectors: Per pixel conversion gain from on-ground characterization for the
  {{Euclid NISP}} instrument,'' in [{\em {{SPIE Astronomical Telescopes}} +
  {{Instrumentation}} 2018}{\nolinebreak\hspace{0.1em}]},   {\bf 10709},
  1070921 (June 2018).

\bibitem{Kubik-2016}
Kubik, B., Barbier, R., and Chabanat, E., ``A {{New Signal Estimator}} from the
  {{NIR Detectors}} of the {{Euclid Mission}},'' {\em Publications of the
  Astronomical Society of the Pacific}~{\bf 128},  104504 (Sept. 2016).

\bibitem{Seshadri-2008}
Seshadri, S., Cole, D.~M., Hancock, B.~R., and Smith, R.~M., ``Mapping
  electrical crosstalk in pixelated sensor arrays,'' in [{\em High {{Energy}},
  {{Optical}}, and {{Infrared Detectors}} for {{Astronomy
  III}}}{\nolinebreak\hspace{0.1em}]},   {\bf 7021},  54--64, SPIE (July 2008).

\bibitem{LeGraet-2022}
Le~Gra{\"e}t, J., Secroun, A., and Barbier, R., ``Euclid {{Near Infrared
  Spectro-Photometer}}: Spatial considerations on {{H2RG}} detectors interpixel
  capacitance and {{IPC}} corrected conversion gain from on-ground
  characterization,'' in [{\em X-{{Ray}}, {{Optical}}, and {{Infrared
  Detectors}} for {{Astronomy X}}}{\nolinebreak\hspace{0.1em}]},   98 (Aug.
  2022).

\bibitem{Finger-2006}
Finger, G., Dorn, R., and Meyer, M., ``Interpixel capacitance in large format
  {{CMOS}} hybrid arrays,'' {\em Proceedings of SPIE - The International
  Society for Optical Engineering}~{\bf 6276} (June 2006).

\bibitem{Kubik-2014a}
Kubik, B., Barbier, R., and Castera, A., ``Impact of noise covariance and
  nonlinearities in {{NIR H2RG}} detectors,'' {\em Proceedings of SPIE - The
  International Society for Optical Engineering}~{\bf 9154},  91541Q (July
  2014).

\bibitem{delaFleche-2023}
{de la Fl{\`e}che}, A.~N., Atteia, J.-L., and Boy, J., ``{{CAGIRE}}: A
  wide-field {{NIR}} imager for the {{COLIBRI}} 1.3 meter robotic telescope,''
  {\em Experimental Astronomy}~{\bf 56},  645--685 (Dec. 2023).

\bibitem{Mosby-2020}
Mosby, G., Rauscher, B.~J., and Bennett, C., ``Properties and characteristics
  of the {{Nancy Grace Roman Space Telescope H4RG-10}} detectors,'' {\em
  Journal of Astronomical Telescopes, Instruments, and Systems}~{\bf 6},
  046001 (Oct. 2020).

\bibitem{Brown-2006}
Brown, M., Schubnell, M., and Tarl{\'e}, G., ``Correlated {{Noise}} and
  {{Gain}} in {{Unfilled}} and {{Epoxy-Underfilled Hybridized HgCdTe
  Detectors}},'' {\em Publications of the Astronomical Society of the
  Pacific}~{\bf 118},  1443--1447 (Oct. 2006).

\end{thebibliography}
\bibliographystyle{spiebib} 

\end{document}